# Quantum Emission from Defects in Single Crystal Hexagonal Boron Nitride

Toan Trong Tran[1], Cameron Zachreson[1], Amanuel Michael Berhane[1], Kerem Bray[1], Russell Guy Sandstrom[1], Lu Hua Li[2], Takashi Taniguchi[3], Kenji Watanabe[3], Igor Aharonovich[1]*, and Milos Toth[1],*

[1]School of Mathematical and Physical Sciences, University of Technology Sydney, Ultimo, NSW, 2007, Australia

[2]Institute of Frontier Materials, Deakin University, Geelong Waurn Ponds Campus, VIC 3216, Australia [3]College of Physics

[3]National Institute for Materials Science, Ibaraki, 305-0044, Japan

*Corresponding author: *Milos.Toth@uts.edu.au*; *Igor.Aharonovich@uts.edu.au*

**Abstract**

*Bulk hexagonal boron nitride (hBN) is a highly nonlinear natural hyperbolic material that attracts major attention in modern nanophotonics applications. However, studies of its optical properties in the visible part of the spectrum and quantum emitters hosted by bulk hBN have not been reported to date. In this work we study the emission properties of hBN crystals in the red spectral range using sub-bandgap optical excitation. Quantum emission from defects is observed at room temperature and characterized in detail. Our results advance the use of hBN in quantum nanophotonics technologies and enhance our fundamental understanding of its optical properties.*

## I. INTRODUCTION

Recent studies have been focused on various wide-band gap materials with superior chemical and thermal stability needed for optoelectronics and applications [1-5]. Bulk hexagonal boron nitride (hBN) is one of these materials and offers excellent thermal conductivity and bright luminescence in the deep ultraviolet (UV) region associated with band-edge transitions [6-11] which is highly advantageous for light emitting devices. Despite numerous studies of these emissions by X-Ray, electron energy loss and luminescence spectroscopy techniques, exact models of the emissions remain under debate. This is in part due to the complicated growth techniques (often high pressure and temperature) that can give rise to significant amounts of dislocations, stacking faults and impurity atoms, including oxygen and carbon. Furthermore, hBN is a naturally hyperbolic material. Its extreme optical anisotropy gives rise to potential interesting applications based on sub-wavelength confinement and strong light-matter interactions [12-14]. However, to fully exploit the potential of hBN for quantum nanophotonics applications, true single photon emitters in this material must be identified.

Here, we report a comprehensive study of single photon emitters in the visible and near infrared regions hosted by bulk hBN crystals. Our results promote the use of hBN as an interesting candidate for emerging applications in quantum technologies and nanophotonics.

## II. EXPERIMENTAL

The hBN single crystals used in this work were produced by the high-temperature and high-pressure process [15]. The hBN crystals were annealed for 30 min at 850°C under 0.5 Torr of Argon in a conventional tube furnace. The samples were heated at a ramp rate of 2°C/sec from room temperature. Upon completion, the samples were cooled to room temperature overnight. The annealing process was used to increase the number of luminescent defect centers. Optical images were obtained using a Zeiss optical microscope. Raman spectroscopy was conducted using a Renishaw in Via Raman™ microscope. The confocal maps and single photon spectroscopy were performed at room temperature using a continuous wave (CW) 532 nm laser (Gem 532™, Laser Quantum Ltd.). The laser was directed through a Glan-Taylor polarizer (Thorlabs Inc.) and a half waveplate, and focused onto the sample using a high numerical aperture (NA = 0.9, Nikon) objective lens. Scanning was performed using an X-Y piezo scanning mirror (FSM-300TM, Newport Corp.) or an X-Y-Z nanocube system (PI instruments). The collected light was filtered using a 532 nm dichroic mirror (532 nm laser BrightLine™, Semrock Inc.) and an additional long pass filter (SemrockTM). The signal was then coupled into a graded index fiber, where the fiber aperture serves as a confocal pinhole. A fiber splitter was used to direct the light into a spectrometer (Acton SpectraPro™, Princeton Instrument Inc.) or into two avalanche photodiodes (Excelitas Technologies™) used for single photon counting. Correlation measurements were done using a time-correlated single photon counting module (PicoHarp300™, PicoQuant™). Lifetime measurements were performed using a 675 nm pulsed laser excitation source (PiL067X™, Advanced Laser Diode Systems™ GmbH) with a 45 ps pulse width.

## III. RESULTS AND DISCUSSION

We start by surveying the sample using conventional optical microscopy and Raman spectroscopy, as shown in Figure 1. The bulk crystal shows a number of visible stacking disorder lines (Figure 1a). Raman spectra of the sample exhibit the characteristic $E_{2g}$ in-plane vibrational mode of bulk hexagonal boron nitride at 1365 cm$^{-1}$, with a full width at half maximum (FWHM) of 8.2 cm$^{-1}$ (Figure 1b), indicating that the sample is a high quality crystal [16]. Additional characterization was performed using near-edge x-ray absorption fine structure (NEXAFS) and cathodoluminescence (CL) spectroscopy. The results are provided in the Supporting Information, along with corresponding data from a second sample that contained a lower concentration of defects (as per NEXAFS and optical microscopy data). We note that the single photon emitters characterized below (Figures 2-4) were not found in this second sample.

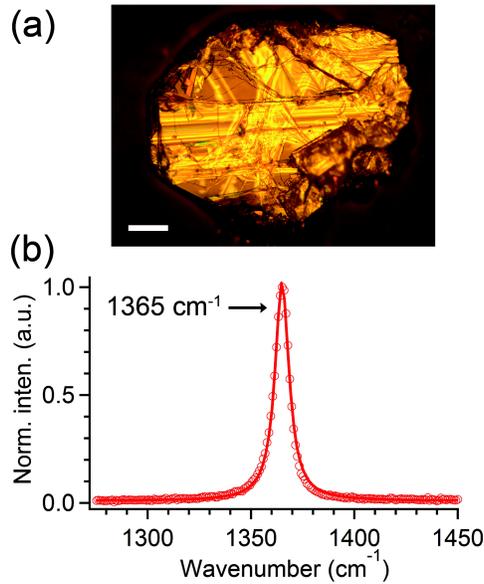

*Fig. 1. (a) Optical microscope image of bulk hBN. The scale bar indicates 100 μm. (b) Raman scattering spectrum obtained with a 633 nm He-Ne laser showing a peak at 1365 cm$^{-1}$ with a FWHM of 8.2 cm$^{-1}$.*

Most prior luminescence studies of bulk hBN were performed using above band-gap excitation (as in Figure S2), whereby the emission spectra are typically dominated by UV or near-UV luminescence. On the contrary, in this letter, we adopt the techniques used for other wide bandgap materials, such as SiC and diamond, in order to access deep, defect-related levels preferentially and avoid transitions that entail the conduction and valence bands. Figure 2a shows a typical confocal photoluminescence (PL) map of the hBN crystal obtained using a 532 nm excitation laser and a shallow confocal depth of a few hundred nanometers, revealing an isolated emission (circled in red) along with other ensemble emissions. A background spectrum (dotted grey trace) obtained from a region adjacent to the emitter is shown in Figure 2b, revealing two Raman lines at 575 nm and 583 nm [17]. A PL spectrum taken from this particular defect (solid red trace in Figure 2b) reveals two sharp peaks at 618 nm and 629 nm. Both peaks are potentially the zero phonon lines (ZPLs) of a color center. Correlation spectroscopy using a Hanbury Brown & Twiss (HBT) interferometer was therefore used to prove that the two peaks correspond to an isolated defect that emits single photons. Figure 2c displays the second order autocorrelation measurement, $g^2(\tau)$, showing that at zero delay time, $g^2(0) \sim 0.35$ (Figure 2c, where the red dots are experimental data while the solid line is a fit obtained using a three level model). The $g^2(0)$ value of less than 0.5 proves unambiguously that the emitter is a single photon source [18-20] (we note that the data are not corrected for background which comprises ~ 30% of the total light intensity). On average, we found two such isolated emitters in each 60 x 60 μm$^2$ area of the sample.

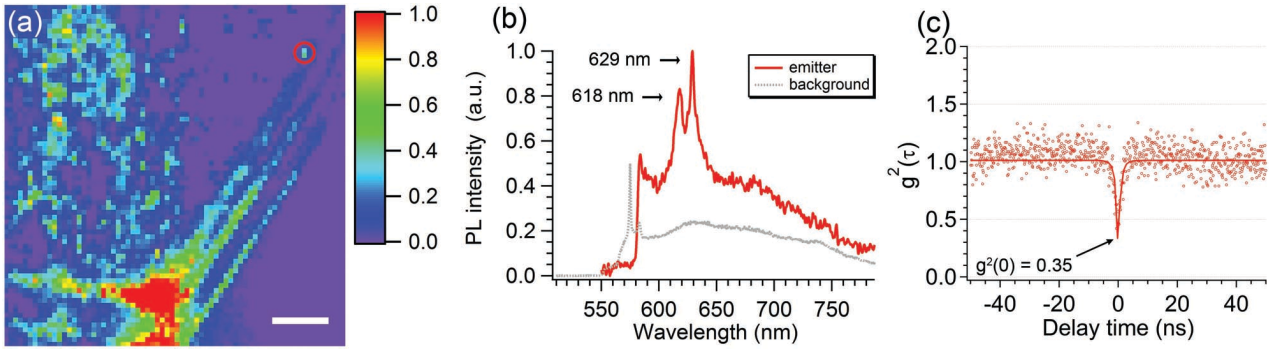

*Fig. 2. Optical characterization performed with a 532 nm continuous wave (CW) laser and a 568 nm long pass filter in the collection pathway. (a) A typical confocal map of bulk hBN showing a number of isolated emission centers and ensembles of these centers. The scale bar indicates 10 μm. (b) A room-temperature photoluminescence spectrum of the isolated emission circled in the PL confocal map. The solid red and dotted grey traces represent emitter and background spectra, respectively. The emitter spectrum reveals a pair of peaks at 618 nm and 629 nm that are potentially the zero-phonon lines of the defect transition. (c) An antibunching curve recorded from the defect center in (b) showing a dip of ~ 0.35, proving the single photon emission nature of the defect.*

The emission associated with the defect characterized in Figure 2 bleached after several minutes of excitation at 532 nm. To gain additional insights into the occurrence of emitters in hBN, we employed a longer excitation wavelength of 675 nm. A typical confocal map is shown in Figure 3a and contains several isolated emission spots. Similar to the previous case, we found an average of two isolated emitters in each 60 x 60 μm$^2$ area of the sample. A PL spectrum (solid red trace) taken from one of these spots (circled in red), shows a broad emission band in the range of ~770 – 900 nm (Figure 3b). The dotted grey trace is the background luminescence [17]. A second order autocorrelation function, g$^2$(τ), is shown in Figure 3c (the red dots are experimental data while the solid line is a fit obtained using a three level model). A dip of ~ 0.37 at zero delay time confirms the single photon emission nature of the center. As our antibunching curve was obtained without background correction, the deviation from zero originates primarily from background emissions from the crystal. The inset of Figure 3c demonstrates triggered single photon emission using pulsed excitation, which is important for many practical nanophotonics applications.

Measurements of fluorescence intensity from the same defect as a function of excitation power show that the fluorescence saturates at a count rate slightly greater than 200,000 counts/sec, comparable with other single photon emitters in bulk materials [21]. Time-resolved fluorescence measurements yield a lifetime of ~1.0 ns (Figure 3d). This value is comparable to most conventional single emitters in bulk materials [21,22].

The electron-phonon coupling characteristics of the two emitters were estimated using the Huang-Rhys (RH) factor. By fitting the PL spectra with multiple Lorentzian peaks, we calculated the RH factor to be 0.93 and 1.93 for the color centers shown in Figure 2 and Figure 3, respectively. These values are in good agreement with prior literature [9].

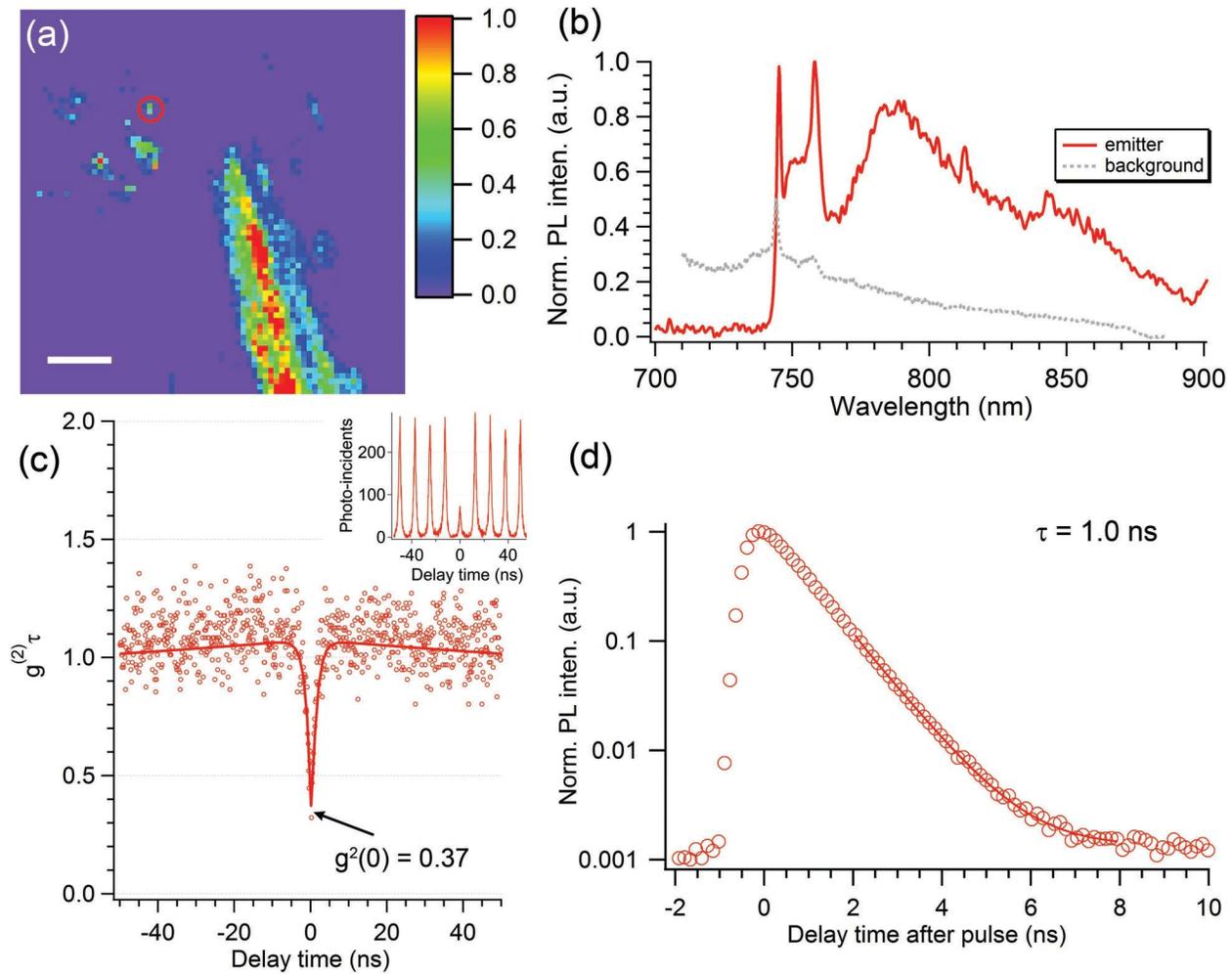

*Fig. 3. (a) A typical confocal map showing isolated emission centers and ensembles of emitters. The scale bar indicates 10 μm. (b) A room-temperature PL spectrum of the isolated emission that is circled in the PL confocal map, revealing a broad emission band at ~ 770 – 900 nm. The solid red and dotted grey traces represent emitter and background spectra, respectively. (c) An antibunching curve recorded by continuous wave excitation of the defect center in (b) showing a dip of ~0.37, proving the quantum nature of the defect. The inset shows a similar antibunching curve obtained by pulsed excitation. (d) Time-resolved fluorescence measurement of the defect center in (b) revealing a very short radiative lifetime of ~ 1.0 ns. All measurements were done using a 675 nm CW laser at room temperature, with a 855 ± 110 nm band pass filter. Pulsed $g^2(\tau)$ and lifetime measurements (d) were conducted using a 675 nm laser with a pulse width of 45 ps, a power of 200 μW, and repetition rate of 80 MHz.*

To characterize the photo dynamics of the defect shown in Figure 3 further, time-correlated measurements were performed on the emitter. The detected photons were time-tagged over a long time scale, and the results are shown in Figure 4a. Three exponential components yield the best fit for the autocorrelation curve. This indicates that there are several additional metastable states associated with the electronic structure of this defect beside its principle radiative transition (i.e. between the ground state and the excited state). An illustration of such an excited electronic structure is shown in the inset of Figure 4a, with lifetimes of the three additional metastable states of 480 ns, 5 μs, and 31 ms, respectively. The transition rates to these states are relatively low, which explains why the defect can be detected on a single photon level with reasonable brightness.

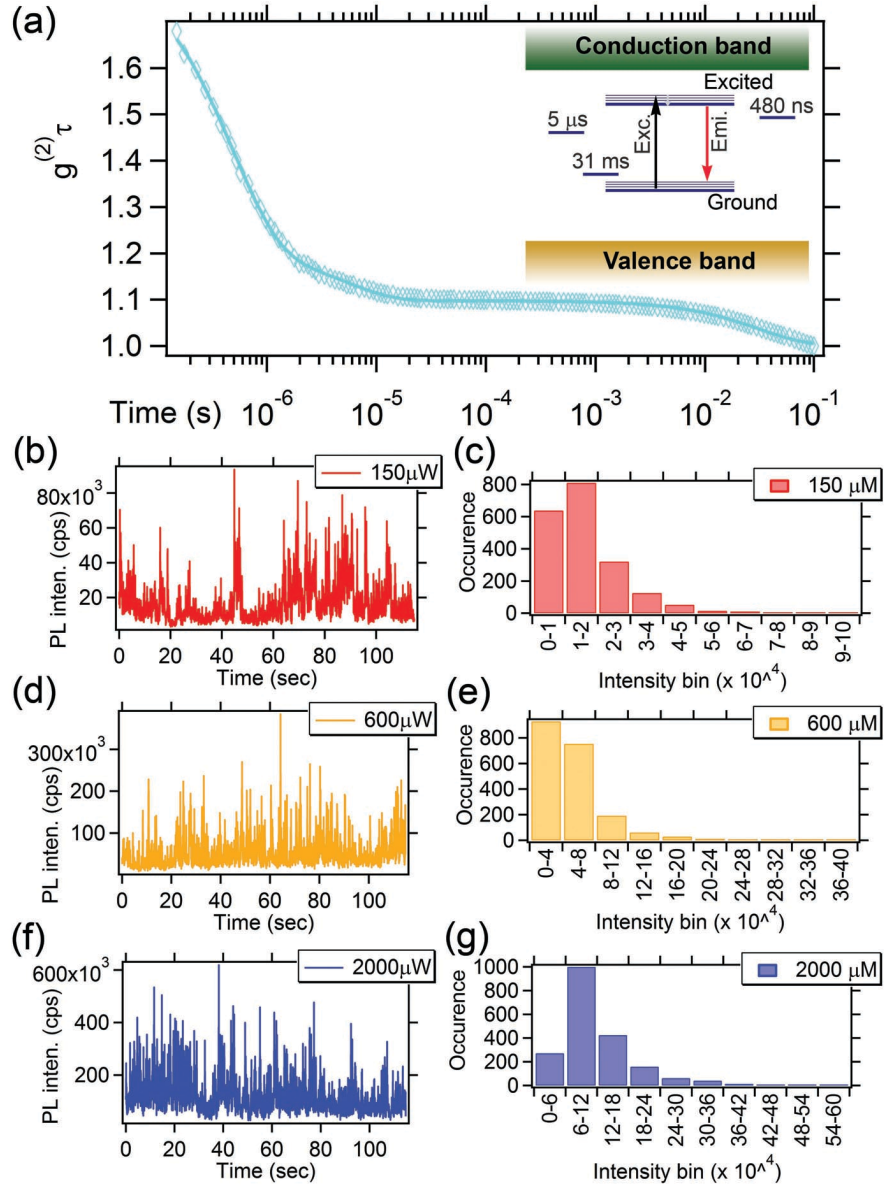

*Fig. 4. (a) Long time-scale second-order autocorrelation function (recorded up to 0.1 s) reveals at least 3 possible metastable states of the defect center characterized in Figure 3. The inset illustrates the possible excited electronic configuration of the defect center, including the existence of three metastable states. Temporal fluorescence intensity fluctuations at (b) 150 µW, (d) 600 µW, and (f) 2000 µW and the corresponding histograms at (c) 150 µW, (e) 600 µW, and (g) 2000 µW.*

To elucidate the complex dynamics further, we characterized the blinking behavior of the defect at elevated excitation power. Plots of fluorescence intensity were recorded as a function of time at excitation powers of 150 µW (Figure 4b), 600 µW (Figure 4d) and 2000 µW (Figure 4f). Blinking is clearly observed at all the investigated excitation powers. However, no bleaching of the defect is observed. This behavior is consistent with the autocorrelation function recorded in Figure 4(a), and confirms the power-dependence of the transition to the metastable states. Histograms of the fluorescence intensity (Figure 4c, 4e, and 4g) reveal asymmetric distributions similar to that observed in ZnO nanoparticles [23]. Such behavior, often referred to as a "photon burst", is characteristic of an intermittent bright state. The severe spectral diffusion is likely associated with close proximity of the emitter to the surface or to other extended defects. Given the wide bandgap of hBN, it is expected that additional defects with other ZPL positions will be found, as in the case of shallow emitters in diamond [24].

## IV. SUMMARY

In summary, room temperature single photon emission from bulk hBN was observed in the visible and the near infrared spectral ranges. The longer wavelength center does not bleach and has a short lifetime of ~1 ns. The defect centers offer a new class of a single photon source to be realized based on bulk hBN. Our results enhance present understanding of fundamental quantum phenomena in hBN and open the door to studies of light-matter interactions with quantum emitters in a hyperbolic medium.

## ACKNOWLEDGMENT

This research was partly undertaken on the soft x-ray beamline at the Australian Synchrotron, Victoria, Australia. The work was supported in part by the Australian Research Council (Project Number DP140102721) and FEI Company. I. A. is the recipient of an Australian Research Council Discovery Early Career Research Award (Project Number DE130100592). Partial funding for this research was provided by the Air Force Office of Scientific Research, United States Air Force.